\documentclass[showpacs,amsmath,amssymb,aps,showkeys,floatfix,prd,a4paper,twocolumn]{revtex4}

\usepackage[dvips]{graphicx}
\usepackage{dcolumn}
\usepackage{bm}
\usepackage{epsfig}
\usepackage{amsfonts}
\usepackage{amssymb,amscd}

\def\lsim{\raise0.3ex\hbox{$<$\kern-0.75em\raise-1.1ex\hbox{$\sim$}}}
\def\gsim{\raise0.3ex\hbox{$>$\kern-0.75em\raise-1.1ex\hbox{$\sim$}}}

\def\beq{\begin{equation}}
\def\eeq{\end{equation}}
\def\bea{\begin{eqnarray}}
\def\eea{\end{eqnarray}}
\def\bq{\begin{quote}}
\def\eq{\end{quote}}

\newcommand{\rr}{\mbox{\boldmath $r$}}

\newcommand{\rb}{\mbox{\boldmath $b$}}

\def\gappeq{\mathrel{\rlap {\raise.5ex\hbox{$>$}}
{\lower.5ex\hbox{$\sim$}}}}

\def\lappeq{\mathrel{\rlap{\raise.5ex\hbox{$<$}}
{\lower.5ex\hbox{$\sim$}}}}

\def\Toprel#1\over#2{\mathrel{\mathop{#2}\limits^{#1}}}

\begin{document}


\title{Exclusive heavy vector meson photoproduction in hadronic collisions at the LHC: predictions of the   
Color Glass Condensate model for Run 2 energies}

\author{V.~P. Gon\c{c}alves $^1$, B. D. Moreira $^2$ and F. S. Navarra $^2$}
\affiliation{$^1$ High and Medium Energy Group, \\
Instituto de F\'{\i}sica e Matem\'atica, Universidade Federal de Pelotas\\
Caixa Postal 354, CEP 96010-900, Pelotas, RS, Brazil \\ $^2$ Instituto de F\'{\i}sica, Universidade de S\~{a}o Paulo,
C.P. 66318,  05315-970 S\~{a}o Paulo, SP, Brazil}

\date{\today}

\begin{abstract}
In this letter we update our predictions for  exclusive $J/\Psi$ and $\Upsilon$ photoproduction   in proton-proton and  nucleus - nucleus  
collisions at the Run 2 LHC energies obtained with the color dipole formalism and considering the impact parameter Color Glass Condensate 
model (bCGC) for  the forward dipole - target scattering amplitude. A comparison with the LHCb data on  rapidity distributions and photon - 
hadron cross sections is presented. Our results demonstrate that the current data can be quite well described by the bCGC model, which takes 
into account  nonlinear effects in the QCD dynamics and reproduces the very precise HERA data, without introducing  any  additional effect 
or free parameter.   
 
\end{abstract}

\maketitle


The study of photon - induced processes in hadronic collisions \cite{upc} became a reality in the last years 
\cite{cdf,star,phenix,alice,alice2,lhcb,lhcb2,lhcb3,lhcbconf} and new data associated to the Run 2 of the LHC are expected to be released soon. 
Theoretically, we expect that these new data allow us to get answers for several important open questions of the Standard Model (SM) as well as to 
shed some light on possible beyond SM physics (For a recent review see \cite{review_forward}).  One these questions is related to the treatment of 
the QCD dynamics at high energies and large nuclei \cite{hdqcd}, which is probed in  exclusive vector meson photoproduction in hadronic collisions 
\cite{bert,vicmag}.  
In the last years, this process was studied by several theoretical groups considering different formalisms and underlying assumptions 
(See e.g. \cite{vicmag_ups,frankfurt_ups,Schafer_ups,Martin}). In particular, in Refs. \cite{bruno,bruno2} we estimated the exclusive $J/\Psi$ and 
$\Upsilon$ photoproduction in hadronic collisions within the dipole formalism considering different models for the vector meson wave functions  and/or 
for the forward dipole - hadron scattering amplitude. Moreover, we have presented a comparison with the Run 1 LHC data and demonstrated that our 
predictions were able  to describe those data if the nonlinear effects in the QCD dynamics are taken into account. Although in Refs. \cite{bruno,bruno2}  
we have presented some predictions for future runs of the LHC, they were calculated for center of mass energies different from those that are being 
considered for the Run 2. One of the motivations for this letter, is to present  predictions that can be directly compared with the expected Run 2 data. 
Another one  is to present for the first time a comparison between our predictions and the  data on the energy dependence of the total 
$\gamma p \rightarrow V p$ ($V = J/\Psi, \, \Upsilon$) cross section, which have been extracted from the data on rapidity distributions of the vector mesons 
photoproduced  in hadronic collisions. Finally, as a by product, we also present a comparison of our prediction \cite{bruno2}  for the exclusive 
$\Upsilon$ photoproduction in $pp$ collisions at $\sqrt{s} = 7$ TeV   with the LHCb data \cite{lhcb3} that have been released after the publication of 
our previous paper.

Initially, let us present a brief review of the main concepts needed to describe the photon -- induced interactions in hadronic collisions and the 
formalism used in our calculations (For a detailed discussion see Refs. \cite{bruno,bruno2}).
The basic idea in  photon-induced processes is that an ultra relativistic charged hadron (proton or nucleus) 
gives rise to strong electromagnetic fields, such that the photon stemming from the electromagnetic field of one of the two colliding hadrons can 
interact with one photon of the other hadron (photon - photon process) or can interact directly with the other hadron (photon - hadron process) \cite{upc}. 
In these processes the total cross section  can be factorized in terms of the equivalent flux of photons into the hadron projectile and the photon-photon 
or photon-target  cross section. In this letter we focus on exclusive vector meson production in photon -- hadron interactions in hadronic collisions. 
The  differential cross section for the production of a  vector meson $V$ at rapidity $Y$  can be expressed as follows:
\begin{widetext}
\begin{eqnarray}
\frac{d\sigma \,\left[h_1 + h_2 \rightarrow   h_1 \otimes V \otimes h_2\right]}{dY}  =  \left[\omega \frac{dN}{d\omega}|_{h_1}\,\sigma_{\gamma h_2 
\rightarrow V \otimes h_2}\left(\omega \right)\right]_{\omega_L} 
 +  \left[\omega \frac{dN}{d\omega}|_{h_2}\,\sigma_{\gamma h_1 \rightarrow V \otimes h_1}\left(\omega \right)\right]_{\omega_R}\,
\label{dsigdy}
\end{eqnarray}
\end{widetext}
where the rapidity ($Y$) of the vector meson in the final state is determined by the photon energy $\omega$ in the collider frame and by the mass $M_{V}$ 
of the vector meson [$y\propto \ln \, ( \omega/M_{V})$]. Moreover, $\sigma_{\gamma h_i \rightarrow V \otimes h_i}$ is the total cross section of 
exclusive vector meson photoproduction, with the symbol
$\otimes$ representing the presence of a rapidity gap in the final state and $\omega_L \, (\propto e^{-y})$ and $\omega_R \, (\propto e^{y})$ denoting  
photons from the $h_1$ and $h_2$ hadrons, respectively. Moreover, $\frac{dN}{d\omega}$ denotes the  equivalent photon 
spectrum  of the relativistic incident hadron, with the flux of a nucleus   being enhanced by a factor $Z^2$ in comparison to the proton one. 
Eq. (\ref{dsigdy}) takes into account the fact that both incident hadrons can be sources of the photons which will interact with the other hadron, with 
the first term on the right-hand side of the Eq. (\ref{dsigdy}) being dominant  at positive rapidities while the second term dominating at negative 
rapidities due to the fact that the photon flux has support at small values of $\omega$, decreasing exponentially at large $\omega$. As in Refs.  
\cite{bruno,bruno2} we will assume that the photon flux associated to the proton and nucleus can be described by  the Dress - Zeppenfeld  \cite{Dress} 
and the relativistic point -- like charge\cite{upc} models, respectively. Additionally,  in our calculations of  exclusive vector meson photoproduction 
in hadronic collisions we will assume that the rapidity gap survival probability $S^2$ (associated to probability of the scattered proton not to dissociate 
due to  secondary interactions) is equal to the  unity. The inclusion of these  absorption effects in $\gamma h$ interactions is still a subject of intense 
debate \cite{Schafer_ups,frankfurt_ups,Martin}.

The main input in Eq. (\ref{dsigdy}) is the $\gamma h \rightarrow V h$  cross section, which  can be written as
\begin{eqnarray}
\sigma (\gamma h \rightarrow V h) & = &  \int_{-\infty}^0 \frac{d\sigma}{d{t}}\, d{t} \nonumber \\  
& = & \frac{1}{16\pi}  \int_{-\infty}^0 |{\cal{A}}^{\gamma h \rightarrow V h }(x,  \Delta)|^2 \, d{t}\,\,,
\label{sctotal_intt}
\end{eqnarray}
with the  amplitude for producing an exclusive vector meson diffractively  being given in the color dipole formalism by
\begin{eqnarray}
 {\cal A}^{\gamma h \rightarrow V h }({x},\Delta)  =  i
\int dz \, d^2\rr \, d^2\rb_h   
 \,\, (\Psi^{V*}\Psi)  \,\,2 {\cal{N}}^h({x},\rr,\rb_h)
\label{amp}
\end{eqnarray}
where  $(\Psi^{V*}\Psi)$ denotes the wave function overlap between the  photon and vector meson wave functions, $\Delta = - \sqrt{t}$ is the momentum 
transfer and $\rb_h$ is the impact parameter of the dipole relative to the hadron target. Moreover, the variables  $\rr$ and $z$ are the dipole transverse 
radius and the momentum fraction of the photon carried by a quark (an antiquark carries then $1-z$), respectively. $ {\cal N}^h (x, \rr, \rb_h)$ is the 
forward dipole-target scattering amplitude (for a dipole at  impact parameter $\rb_h$) which encodes all the information about the hadronic scattering, 
and thus about the nonlinear and quantum effects in the hadron wave function. It  depends on the $\gamma h$  center - of - mass reaction energy, 
$W = [2 \omega \sqrt{s}]^{1/2}$, through the variable $ x = m^2_V/W^2$. As in Refs. \cite{bruno,bruno2}, in what follows we will consider the Boosted 
Gaussian model \cite{KT,KMW} for the overlap function and the impact parameter Color Glass Condensate (bCGC) model \cite{KMW} for the dipole -- proton 
scattering amplitude ${\cal{N}}^p$. As demonstrated in Ref. \cite{amir},  these models allow us to successfully describe  the high precision combined 
HERA data on inclusive and exclusive processes. In the case of a nuclear target, we will assume that the forward dipole-nucleus amplitude can be expressed 
as follows
\begin{eqnarray}
{\cal{N}}^A(x,\rr,\rb_A) = 1 - \exp \left[-\frac{1}{2}  \, \sigma_{dp}(x,\rr^2) 
\,A\,T_A(\rb_A)\right] \,\,,
\label{enenuc}
\end{eqnarray}
where  $T_A(\rb_A)$ is  the nuclear profile function, which is obtained from a 3-parameter Fermi distribution for the nuclear
density normalized to $1$, and the  $\sigma_{dp}$ is the dipole-proton cross section is expressed by
\begin{eqnarray}
\sigma_{dp} = 2 \,\int d^2\rb_p \, {\cal{N}}^p({x},\rr,\rb_p)
\end{eqnarray}
with ${\cal{N}}^p$ given by the bCGC model. Finally, as in Refs. \cite{bruno,bruno2}, we also include in our calculations the corrections associated to 
the real part of the amplitude and the skewness factor, which is related to the fact that the gluons attached to the $q\bar{q}$ pair can carry different 
light-cone momentum fractions $x$, $x^{\prime}$ of the target. 

\begin{figure}
\includegraphics[scale=0.35]{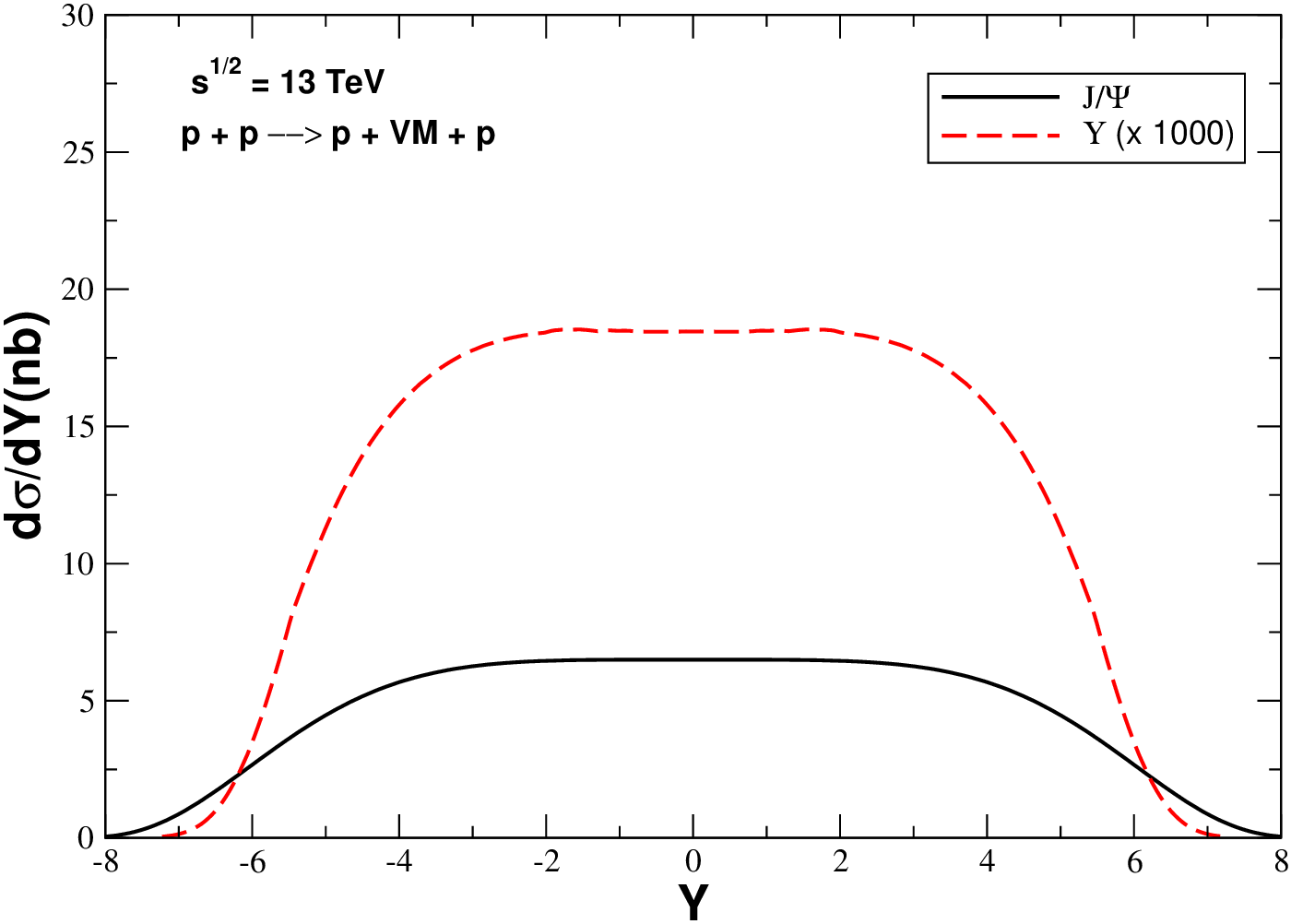}
\includegraphics[scale=0.35]{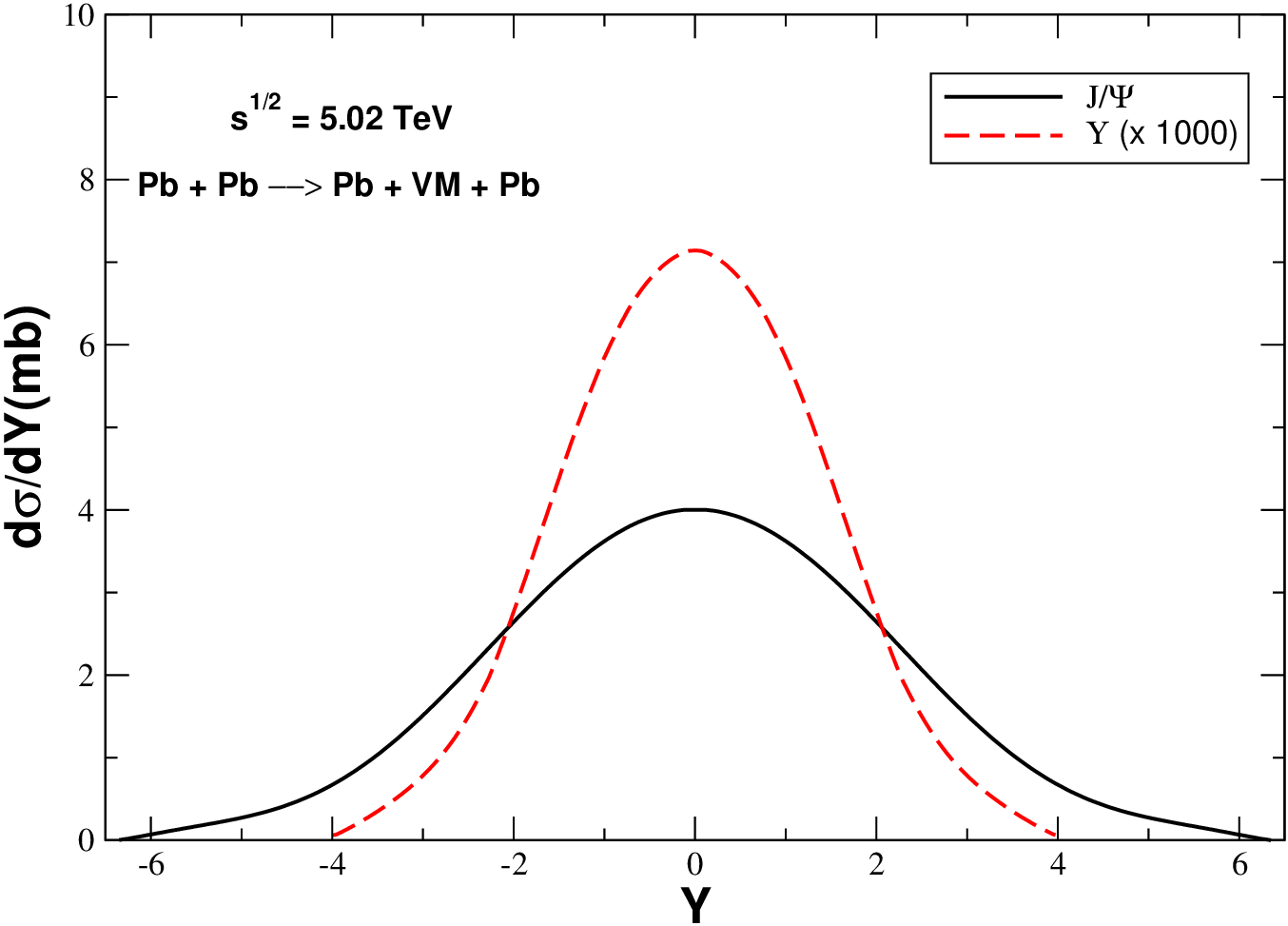} 
\caption{Rapidity distributions for the exclusive $J/\Psi$ and $\Upsilon$ photoproduction in $pp$ (left panel) and $PbPb$ (right panel) collisions.}
\label{fig1}
\end{figure}

\begin{table}[t] 
\centering
\begin{tabular}{||c|c|c||} 
\hline 
\hline
       &  $J/\Psi$ & $\Upsilon$ \\ 
\hline
\hline
$pp$ ($\sqrt{s} = $ 13 TeV) & 72.4 nb & 189.5 pb \\
\hline
$Pb Pb$  ($\sqrt{s} = $ 5.02 TeV) & 21.6 mb & 26.02 $\mu$b \\
\hline
\hline
\end{tabular}
\caption{Total cross sections for the exclusive $J/\Psi$ and $\Upsilon$ photoproduction in $pp$ and $PbPb$ collisions at the Run 2 LHC energies.}  
\label{tab1}
\end{table}

In Fig. \ref{fig1} we present our predictions for the rapidity distributions for  exclusive $J/\Psi$ and $\Upsilon$ photoproduction in $pp$ (upper panel) and 
$PbPb$ (lower panel) collisions considering the center of mass energies of the Run 2. The values of the corresponding cross sections are show in Table 
\ref{tab1}. In comparison with our previous predictions \cite{bruno,bruno2} we observe that the values are smaller by $\approx 10 \%$. It is important to 
emphasize that the bCGC predictions for the $\Upsilon$ production in $PbPb$ collisions are being presented for the first time, since in our previous paper 
\cite{bruno2} this scenario was not considered.

\begin{figure}[t]
\includegraphics[scale=0.35]{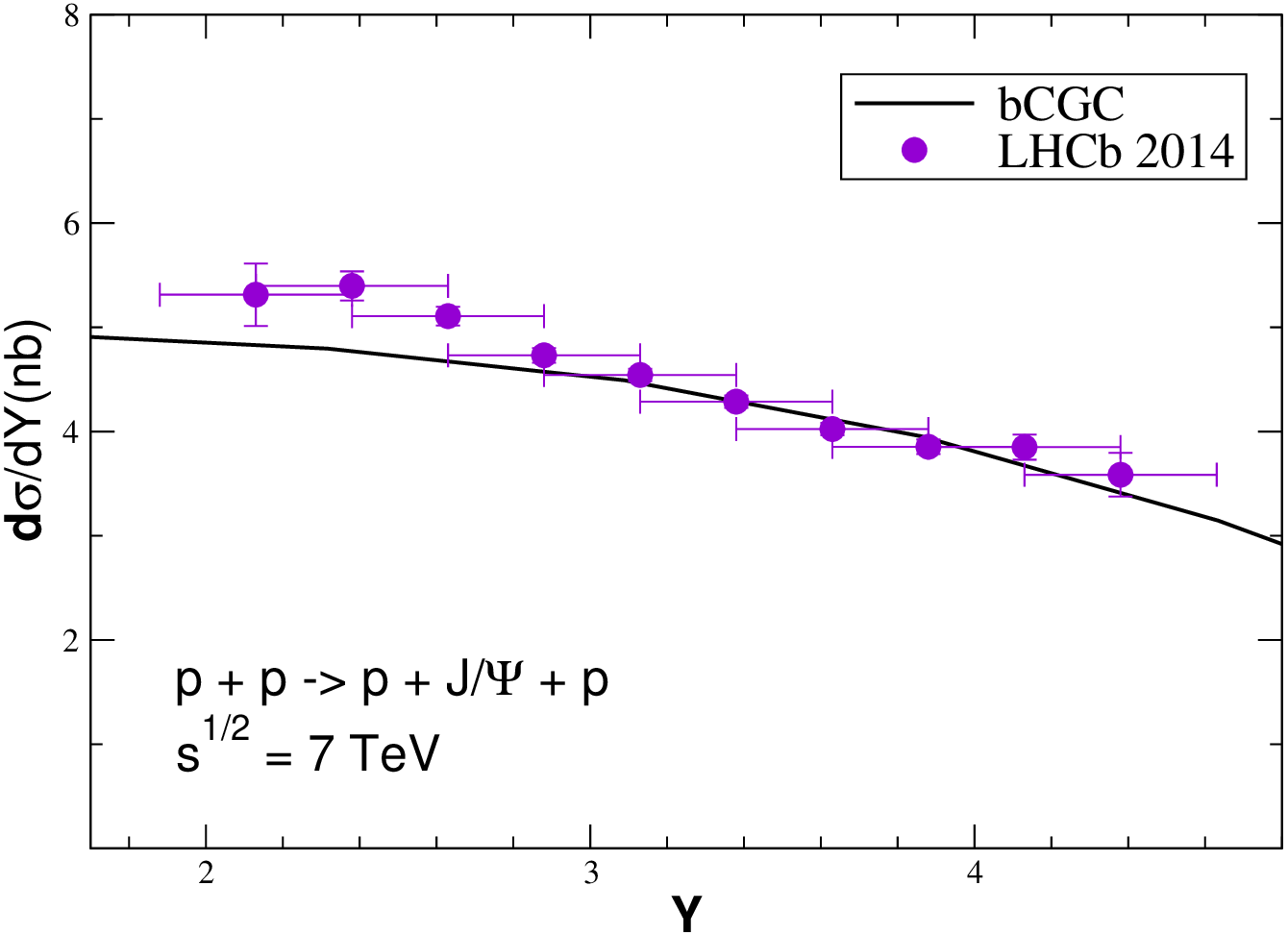}  
\includegraphics[scale=0.35]{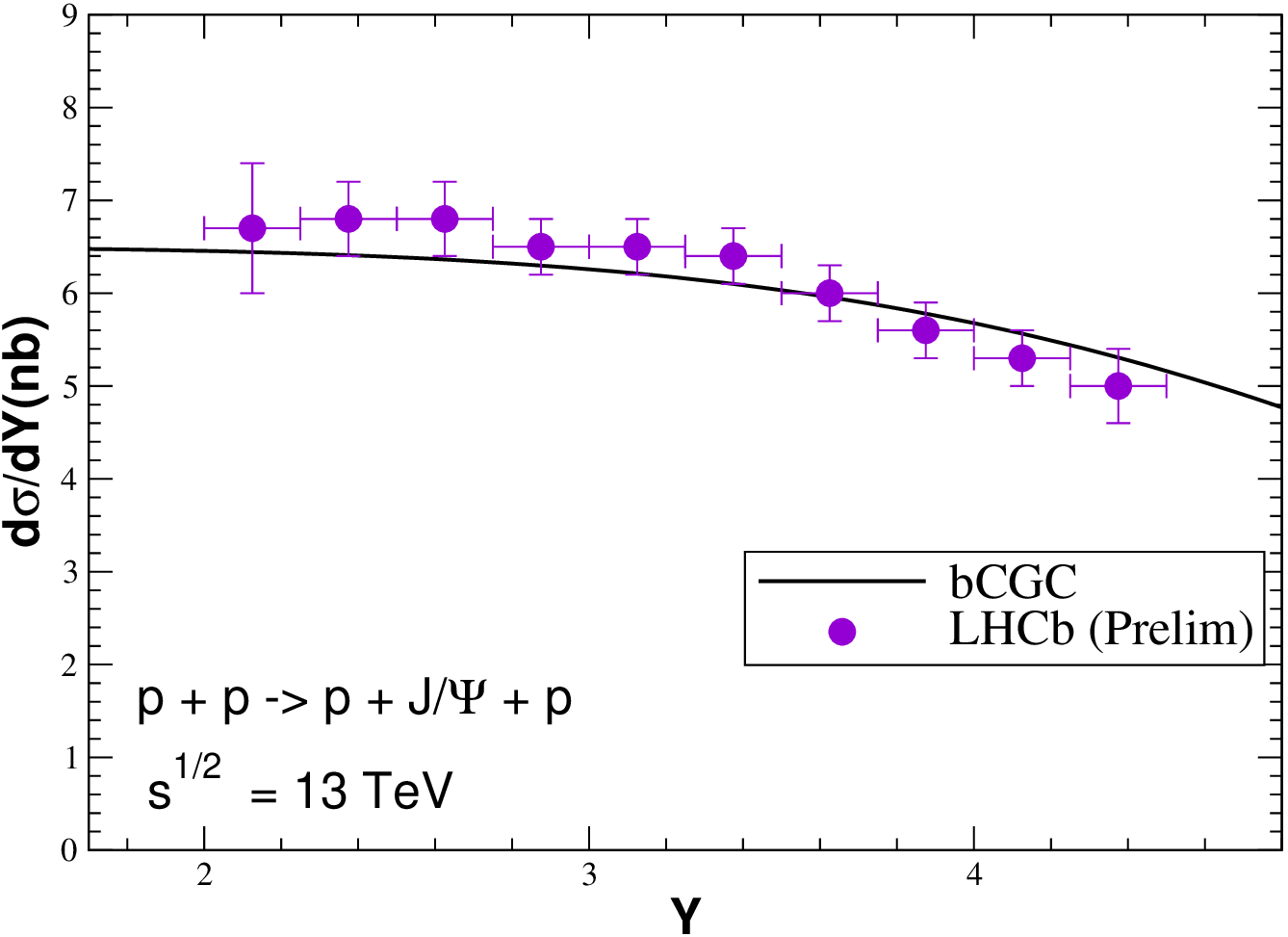}
\caption{Rapidity distributions for the exclusive $J/\Psi$  photoproduction in $pp$  collisions at $\sqrt{s} = 7$ (upper panel) and 13 TeV (lower panel). 
Data from LHCb Collaboration \cite{lhcb2,lhcbconf}. }
\label{fig2}
\end{figure}

\begin{figure}[t]
\includegraphics[scale=0.35]{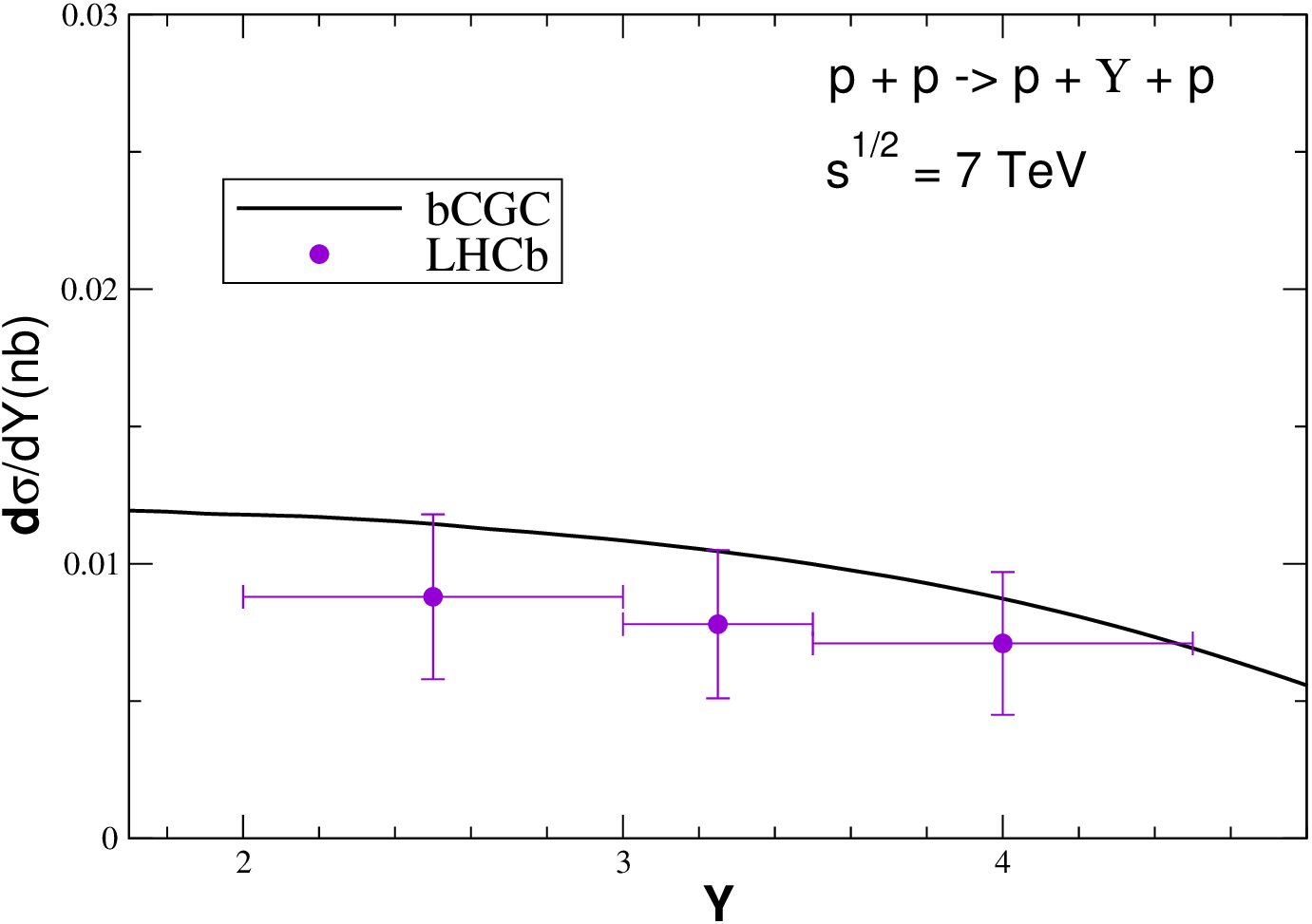}  
\includegraphics[scale=0.35]{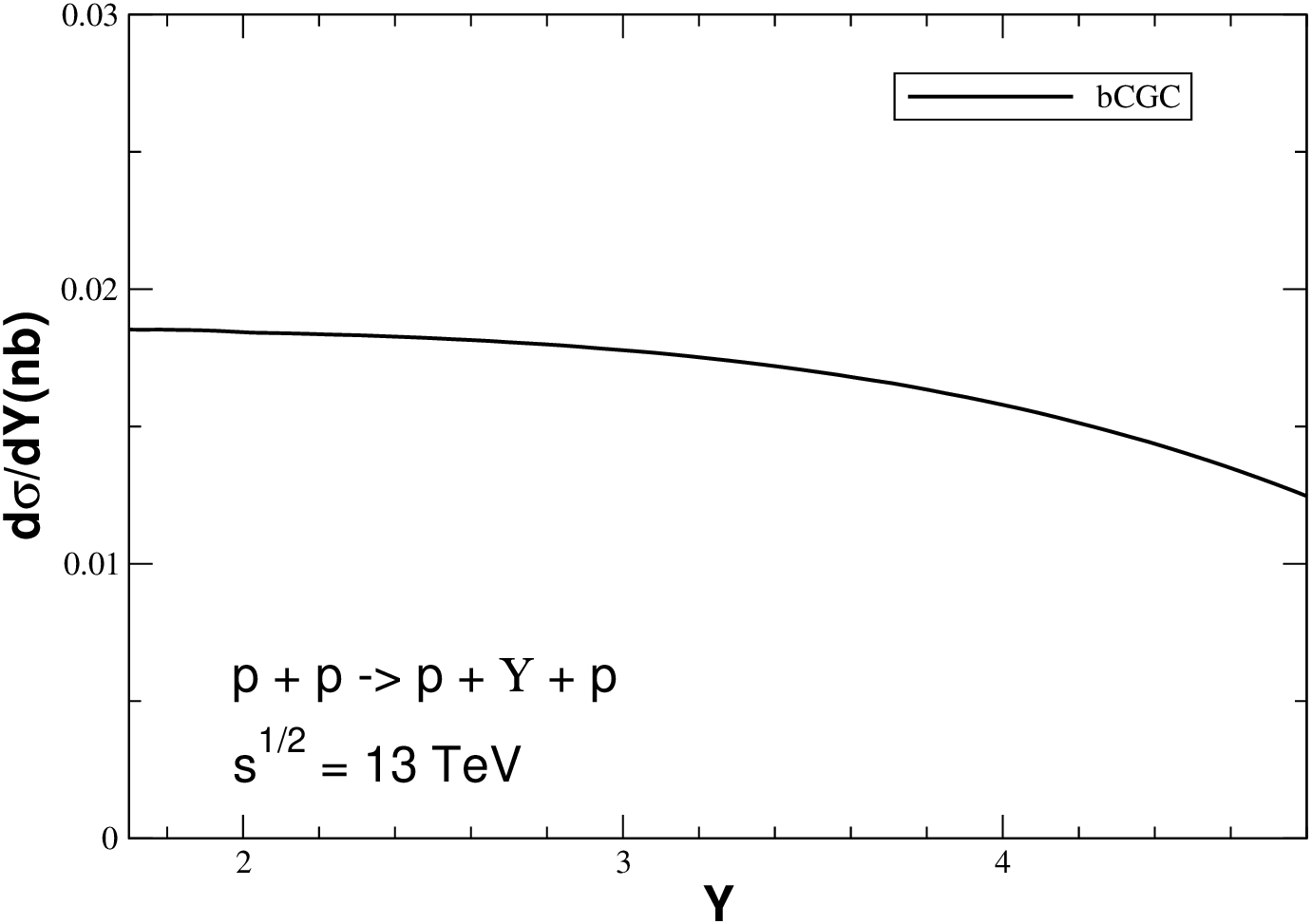}
\caption{Rapidity distributions for the exclusive $\Upsilon$  photoproduction in $pp$  collisions at $\sqrt{s} = 7$ (upper panel) and 
13 TeV (lower panel). Data from LHCb Collaboration \cite{lhcb3}. }
\label{fig3}
\end{figure}

In what follows we will concentrate our analysis on  exclusive vector meson photoproduction in  $pp$ collisions, which was studied by the LHCb 
Collaboration and allows us to do a more detailed comparison of our predictions with the experimental data. In particular, the LHCb Collaboration has 
recently released \cite{lhcbconf} the first (preliminary) data on exclusive $J/\Psi$ production at $\sqrt{s} = 13$ TeV. In Fig. \ref{fig2} we compare 
our predictions for the rapidity distributions measured in exclusive $J/\Psi$  photoproduction in $pp$  collisions at $\sqrt{s} = 7$ (upper panel) and 
13 TeV (lower panel) with the corresponding LHCb data \cite{lhcb2,lhcbconf}. We observe that the bCGC predictions describe these data quite well, without 
the need of modifying the original parameters of the model or introducing any additional physical effect. In  Fig. \ref{fig3} (upper panel) we present 
our predictions for  exclusive $\Upsilon$ production and compare them with the LHCb data for $\sqrt{s} = 7$ TeV \cite{lhcb3}. We can see that also for this 
final state, the bCGC model successfully describes the data. Our prediction for $\sqrt{s} = 13$ TeV, which is expected to be reached in the Run 2,  is also 
presented (lower panel). The comparison of our prediction with future experimental data will be an important check of the bCGC model, since  $\Upsilon$  
production probes smaller dipole separations in comparison to $J/\Psi$ production. While the $J/\Psi$ photoproduction is expected to probe the nonlinear 
regime of the QCD dynamics, the $\Upsilon$ should be sensitive to the transition between the linear and nonlinear regimes. Therefore, we believe that a 
unified description of the exclusive $J/\Psi$ and $\Upsilon$ photoproduction is one important test of the underlying dynamics.

\begin{figure}[t!]
\includegraphics[scale=0.35]{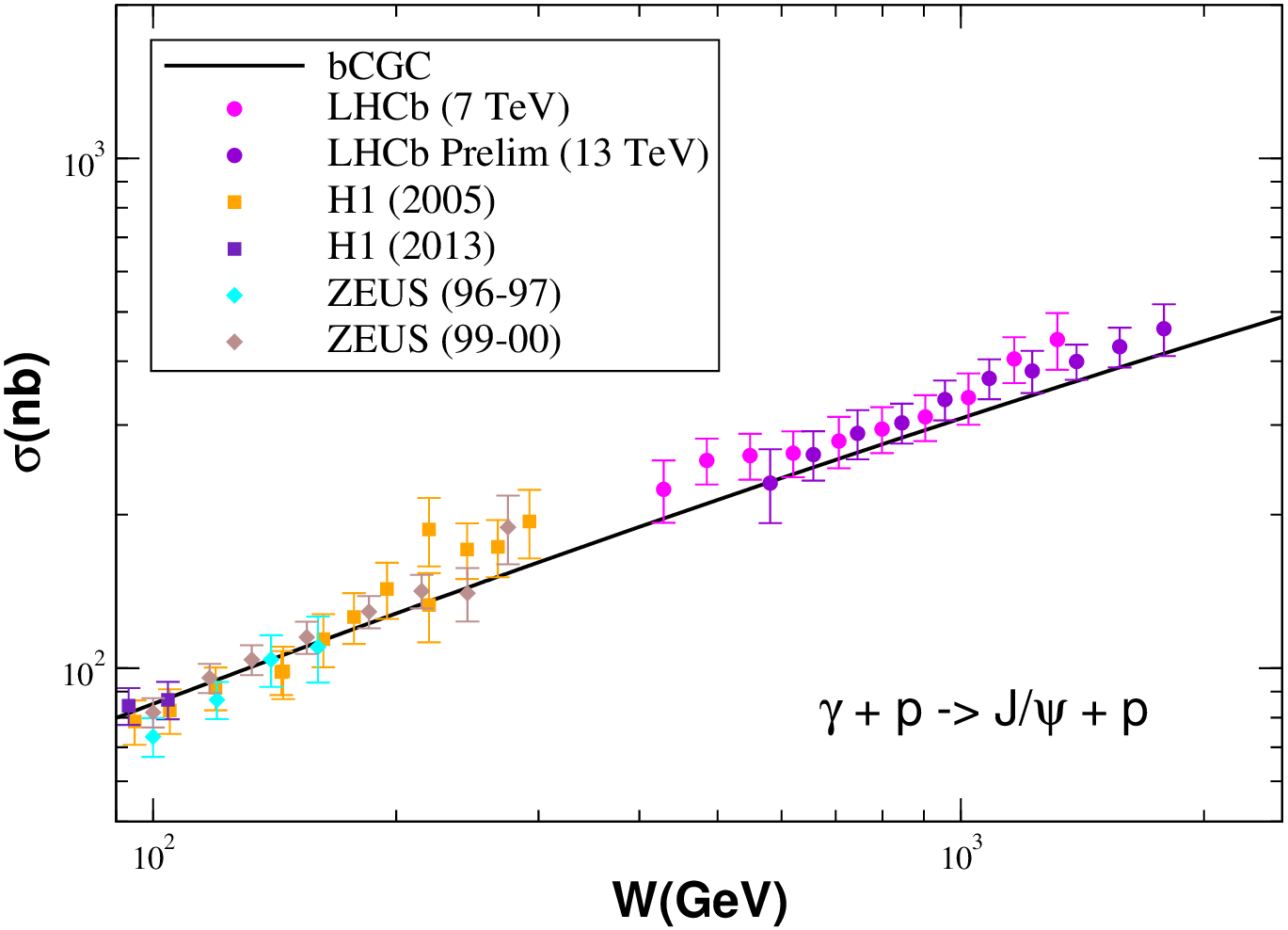} 
\includegraphics[scale=0.35]{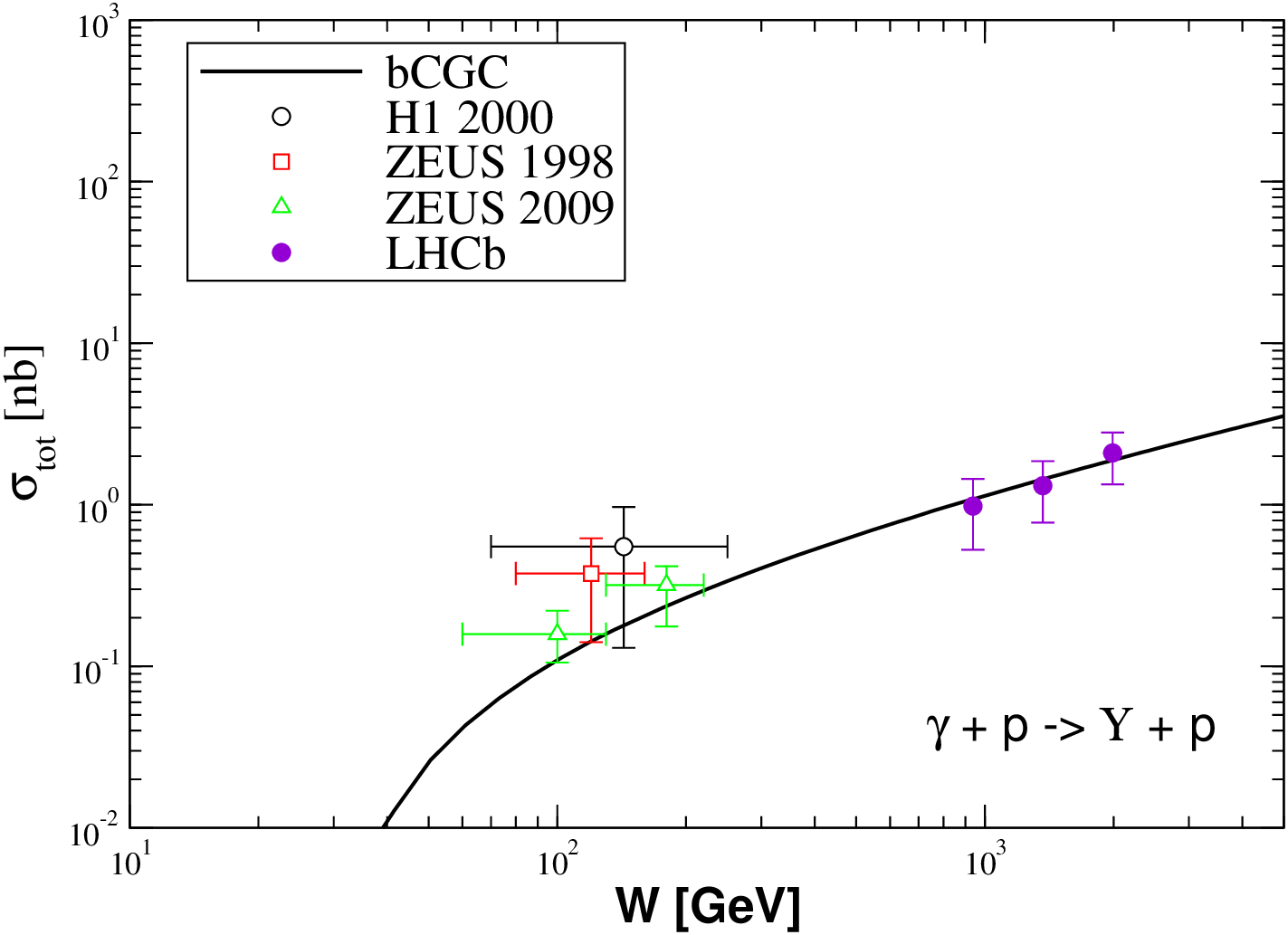}
\caption{Energy dependence of the exclusive $J/\Psi$ (upper panel) and $\Upsilon$ (lower panel) $\gamma p$ cross sections. 
Data from HERA \cite{hera,hera_data_ups} and LHCb \cite{lhcb2,lhcbconf,lhcb3}. }
\label{fig4}
\end{figure}

\hspace{2cm}

One of the main motivations to study the rapidity distributions is that they allow us to access the energy dependence of the $\gamma h \rightarrow V h$ 
cross sections in a new kinematical range, which was not probed e.g. in $ep$ collisions at HERA. The presence of nonlinear effects in the QCD dynamics 
is predicted to modify the energy behavior of the cross sections, since  the growth of the energy implies that smaller values of $x \approx M_V^2/W^2$ 
are probed in the  forward dipole -- target scattering amplitude. Moreover, due to the difference of masses between the $J/\Psi$ and $\Upsilon$, the studies  
of both mesons are complementary. In  Fig. \ref{fig4} we compare our predictions with the LHCb data derived following the procedure presented in 
Refs. \cite{lhcb3,lhcbconf}. In particular, in the upper panel  we can see that our predictions describe quite well the preliminary LHCb data on  the 
exclusive $J/\Psi$ photoproduction in $pp$ collisions at $\sqrt{s} = 13$ TeV, which cannot be described by a simple power - law fit of the HERA data 
\cite{lhcbconf}. Similar agreement  is also observed in the case of the $\Upsilon$ production, shown in the lower panel of Fig. \ref{fig4}. These conclusions 
are not unexpected, since the bCGC model describes the data for the rapidity distributions.

\hspace{2cm}

Finally, let us summarize our main conclusions. Recent experimental results have demonstrated that the study of  hadronic physics using photon induced 
interactions in $pp/pA/AA$ colliders is feasible. In particular,  $\gamma h$ interactions at LHC  probe a kinematical range unexplored by previous 
colliders. The outcoming data on  exclusive $J/\Psi$ and $\Upsilon$ photoproduction in hadronic collisions probe a kinematical range where nonlinear 
effects are expected to strongly affect the QCD dynamics. In our previous studies, we have shown that using the dipole framework and taking into account  
saturation effects (as  in the bCGC model) we are  able to describe the Run 1 LHC data.  In this letter we have updated our comparison with the Run 1 LHCb data 
and we  present our predictions for the energies considered in the Run 2. Our results demonstrated that the bCGC model reproduces  the Run 1 data as well 
as the preliminary data on  $pp$ collisions at $\sqrt{s} = 13$ TeV. Moreover, we have shown that the model also is able to describe the current data on 
the $\gamma p \rightarrow V p$ cross section. Considering that the bCGC model is also  able to describe the inclusive and exclusive HERA data, these results 
suggest that in order to understand $\gamma h$ interactions at high energies we need to take into account QCD  nonlinear effects. 
We strongly believe that a comprehensive analysis of the experimental data on  exclusive light and heavy vector meson photoproduction in hadronic collisions 
at the Run 2 will allow to discriminate between the different approaches to the QCD dynamics.


\section*{Acknowledgements}
 This work was partially financed by the Brazilian funding agencies CAPES, CNPq,  FAPESP and FAPERGS.



\end{document}